\documentclass[review]{elsarticle}
\usepackage{lineno,hyperref}
\usepackage{amsmath}

\usepackage{graphicx}
\usepackage{subfig}
\usepackage{float}
\usepackage{multirow}
\usepackage{natbib}
\usepackage{xcolor}
\setcitestyle{numbers}
\setcitestyle{square}
\floatstyle{plaintop}
\restylefloat{table}

\modulolinenumbers[5]

\journal{journal}

\biboptions{authoryear}

\begin{document}
	
	\begin{frontmatter}
		
		\title{
			An Ensemble Machine Learning Approach for Screening Covid-19 based on Urine Parameters}
		
		\author[1]{Behzad Moayedi}
		\ead{moayedi@aut.ac.ir}
		
		\author[2]{Abdalsamad Keramatfar}
		\ead{keramatfar@rcdat.ir}
		
		\author[1]{Mohammad Hadi Goldani\corref{cor2}}
		\ead{goldani@aut.ac.ir}
		
		\author[3]{Mohammad Javad Fallahi}
		\ead{fallahimj@sums.ac.ir}
		
		\author[4]{Alborz Jahangirisisakht}
		\ead{jahangirsisakht@yahoo.com}
		
		\author[3]{Mohammad Saboori}
		\ead{saboori@sums.ac.ir}
		
		\author[3]{Leyla badiei}
		\ead{badiei@sums.ac.ir}

		\address[1]{Computer Engineering Department, Amirkabir University of Technology, Tehran,Iran}
		\address[2]{Research Center for Development of Advanced Technologies, Tehran, iran}
		\address[3]{Shiraz University of Medical Sciences, Shiraz,Iran}
		\address[4]{Yasuj University of Medical Sciences, Yasuj, Iran}

		\cortext[cor2]{Corresponding author}
		
		\begin{abstract}
			
The rapid spread of COVID-19 and the emergence of new variants underscore the importance of effective screening measures. Rapid diagnosis and subsequent quarantine of infected individuals can prevent further spread of the virus in society. While PCR tests are the gold standard for COVID-19 diagnosis, they are costly and time-consuming. In contrast, urine test strips are an inexpensive, non-invasive, and rapidly obtainable screening method that can provide important information about a patient's health status. In this study, we collected a new dataset and used the RGB (Red Green Blue) color space of urine test strips parameters to detect the health status of individuals. To improve the accuracy of our model, we converted the RGB space to 10 additional color spaces. After evaluating four different machine learning models, we proposed a new ensemble model based on a multi-layer perceptron neural network. Although the initial results were not strong, we were able to improve the model's screening performance for COVID-19 by removing uncertain regions of the model space. Ultimately, our model achieved a screening accuracy of 80\% based on urine parameters. Our results suggest that urine test strips can be a useful tool for COVID-19 screening, particularly in resource-constrained settings where PCR testing may not be feasible. Further research is needed to validate our findings and explore the potential role of urine test strips in COVID-19 diagnosis and management.
		\end{abstract}

			\end{frontmatter}
	
	
	\section{Introduction}
	\label{sec:intro}
	
	Severe acute respiratory syndrome coronavirus type 2 (SARS-CoV-2), which causes coronavirus disease 2019 (COVID-19), has infected more than 550 million people and caused more than 6.3 million deaths worldwide (as of July 6, 2022, according to WHO data) \citep{WHO:2022}. Identification of risk factors and prognostic markers for the development of COVID-19 disease and its consequences is urgently needed to enable early identification and follow-up of high-risk patients. In this regard, researchers around the world tried to find diagnosis methods based on different indicators of health including chest scan and voice quality\citep{roberts2021common,kovacs2021sensitivity,asiaee2020voice,despotovic2021detection}
	
	Recently, biomarker studies have been encouraged around the world to diagnose COVID-19 at an early stage, aiming to target treatment more quickly and reduce health problems associated with the disease\Citep{banerjee2020use,wu2020rapid}. 
	
	A dipstick or urine test strip is an indispensable tool in medical diagnostics to gain a rapid survey of the state of health of the patient\citep{yetisen2013based}. A standard urine test strip can consist of at least ten different buffers or chemical reagents that react (with color change) when immersed in and then removed from a urine sample. After immersion, the test can often be read in as little as 60 to 120 seconds, although specific tests require a longer time. Routine urinalysis with multiparameter strips is the first step in diagnosing a wide range of diseases. The analysis includes tests for the presence of proteins, glucose, ketones, hemoglobin, bilirubin, urobilinogen, acetone, ascorbic acid, nitrite, and leukocytes and tests for pH and specific gravity or tests for infection with different pathogens\citep{strasinger2010anlisis}.
	Due to the ease of performing the urine test and its non-invasiveness, anyone without the need for special instructions can use it.
	\cite{bonetti2020urinalysis,chen2020deep,de2021diagnosis,liu2020value,erdogan2021there} showed that urinalysis parameters can be used for predicting severity and detection of coronavirus disease 2019 (COVID-19). Due to the availability, simplicity, and the low cost of preparing urine strips (\$55 for a pack contains 100 strips) and high price (it costs between \$100 and \$200)\footnote{https://driphydration.com/blog/rapid-pcr-test-cost/}
	 and discomfort of PCR, carrying out this plan can help to quickly diagnose the disease and prevent the spread of the disease.
	
	Machine learning (ML) is a powerful and innovative tool that can help healthcare professionals, policy makers, and other stakeholders during decision-making processes. These computer system models learn and adapt information using algorithms and statistical networks to analyze and draw conclusions from patterns in data\citep{de2021diagnosis}.

	This paper aims to test the ability to use urine parameters to detect Covid-19. Then we try to propose and develop a system based on machine learning that can detect patients with corona disease based on their urine tests. For this purpose, a new dataset has been collected, and several machine learning models have been tested after studying the statistical relation of parameters. Finally, based on the best model, a new ensemble model proposed.

	The rest of the paper is organized as follows: Section \ref{sec:Section 2} introduces the material and methods of this paper. Section \ref{sec:Section 3} presents the experimental results in this paper. Section \ref{sec:Section 5} concludes this manuscript and indicates future works.

	\section{Material and methods}
	\label{sec:Section 2}
	
	\subsection{Data set}
\begin{figure}[H]
		\centering
		\includegraphics[scale=0.6]{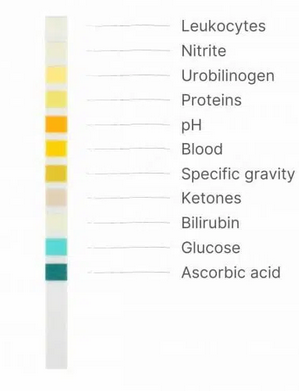}
		\caption{A sample of urine test strip and its buffers}
		\label{fig:Urine_test}
	\end{figure}

	In order to provide data for the training of the models, samples of suspected corona patients were collected from 4 Covid-19 test centers in the cities of Yasuj, Shiraz, and Tehran. The method of collection was that a urine sample and clinical information were received from each of the clients for PCR testing, and in a short period after receiving the urine sample, an image was captured from the urine test strip (the Combi 11 made by Machey Nagel). Figure \ref{fig:Urine_test} shows a urine test strip. The process of data collection includes two stages: gathering and analyzing. Information of this dataset shows in Table \ref{tab:1}.
	
	\begin{table}[H]
		\centering
		\begin{tabular}{|c|c|}
			\hline
			\textbf{Statistics}                     & \textbf{Value}           \\ \hline
			Number of Samples                        & 5293                     \\ \hline
			Average of age                          & 38.41                    \\ \hline
			Gender distribution                     & Male: 3023 - Female: 2270 \\ \hline
			Distribution of healthy and sick people & Healthy: 3304 - Sick: 1989 \\ \hline
		\end{tabular}
		\caption{Statistics of collected data sets}
		\label{tab:1}
	\end{table}

	\subsubsection{Data Gathering}
	The data relating to people who were referred to 4 corona test centers in Yasuj, Shiraz, and Tehran have been collected. The gender information of these clients by the laboratory is shown in Table \ref{tab:10}. Also, Table\ref{tab:11} provides clinical information on patients with respect to three health indicators: diabetes, blood pressure, and smoking. The table presents the number of positive and negative cases of each indicator for each of the four data centers (A, B, C, and D).
	
	For diabetes, data center B has the highest number of positive cases (112), followed by data center A (39), data center D (26), and data center C (9). Similarly, for blood pressure, data center B has the highest number of positive cases (153), followed by data center A (44), data center D (39), and data center C (15).
	
	Regarding smoking, data center B has the highest number of positive cases (263), followed by data center D (42), data center A (13), and data center C (0). It is worth noting that data center C has no positive cases of smoking.
	
	These results provide valuable insights into the prevalence of these health indicators across different data centers. They can be used to identify areas where more attention is needed for disease prevention and management. Further analysis could be conducted to investigate the potential factors that contribute to the differences observed among the data centers.

	\begin{table}[H]
		\centering
		\begin{tabular}{|c|c|c|}
			\hline
			\textbf{Data Center} & \textbf{Male} & \textbf{Female} \\ \hline
			A                    & 427           & 327             \\ \hline
			B                    & 1694          & 1230            \\ \hline
			C                    & 244           & 215             \\ \hline
			D                    & 254           & 311             \\ \hline
		\end{tabular}
		\caption{Gender of clients by different laboratories}
		\label{tab:10}
	\end{table}
	
	\begin{table}[H]
		\centering
		\begin{tabular}{|c|cc|cc|cc|}
			\hline
			\textbf{Statistics}  & \multicolumn{2}{c|}{\textbf{Diabets}}                      & \multicolumn{2}{c|}{\textbf{Blood pressure}}               & \multicolumn{2}{c|}{\textbf{Smoking}}                      \\ \hline
			\textbf{Data center} & \multicolumn{1}{c|}{\textbf{Positive}} & \textbf{Negative} & \multicolumn{1}{c|}{\textbf{Positive}} & \textbf{Negative} & \multicolumn{1}{c|}{\textbf{Positive}} & \textbf{Negative} \\ \hline
			A                    & \multicolumn{1}{c|}{39}                & 715               & \multicolumn{1}{c|}{44}                & 710               & \multicolumn{1}{c|}{13}                & 741               \\ \hline
			B                    & \multicolumn{1}{c|}{112}               & 2812              & \multicolumn{1}{c|}{153}               & 2771              & \multicolumn{1}{c|}{263}               & 2661              \\ \hline
			C                    & \multicolumn{1}{c|}{9}                 & 450               & \multicolumn{1}{c|}{15}                & 444               & \multicolumn{1}{c|}{0}                 & 459               \\ \hline
			D                    & \multicolumn{1}{c|}{26}                & 538               & \multicolumn{1}{c|}{39}                & 525               & \multicolumn{1}{c|}{42}                & 522               \\ \hline
		\end{tabular}
		\caption{The clinical information of the patients according to different laboratories}
		\label{tab:11}
	\end{table}
	
	\subsubsection{Data analysis}
\begin{figure}[H]
		\centering
		\includegraphics[scale=0.6]{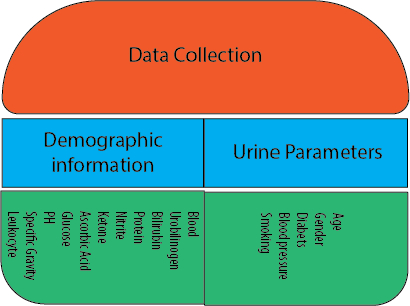}
		\caption{List of parameters of each sample}
		\label{fig:analysis}
	\end{figure}

	For analyzing the images that were captured from the urine test strip, all of the images of a sample are converted to red, green, and blue. In addition to this information, Figure\ref{fig:analysis} Shows list of parameters of each sample.

\begin{figure}[H]
		\centering
		\includegraphics[scale=0.3]{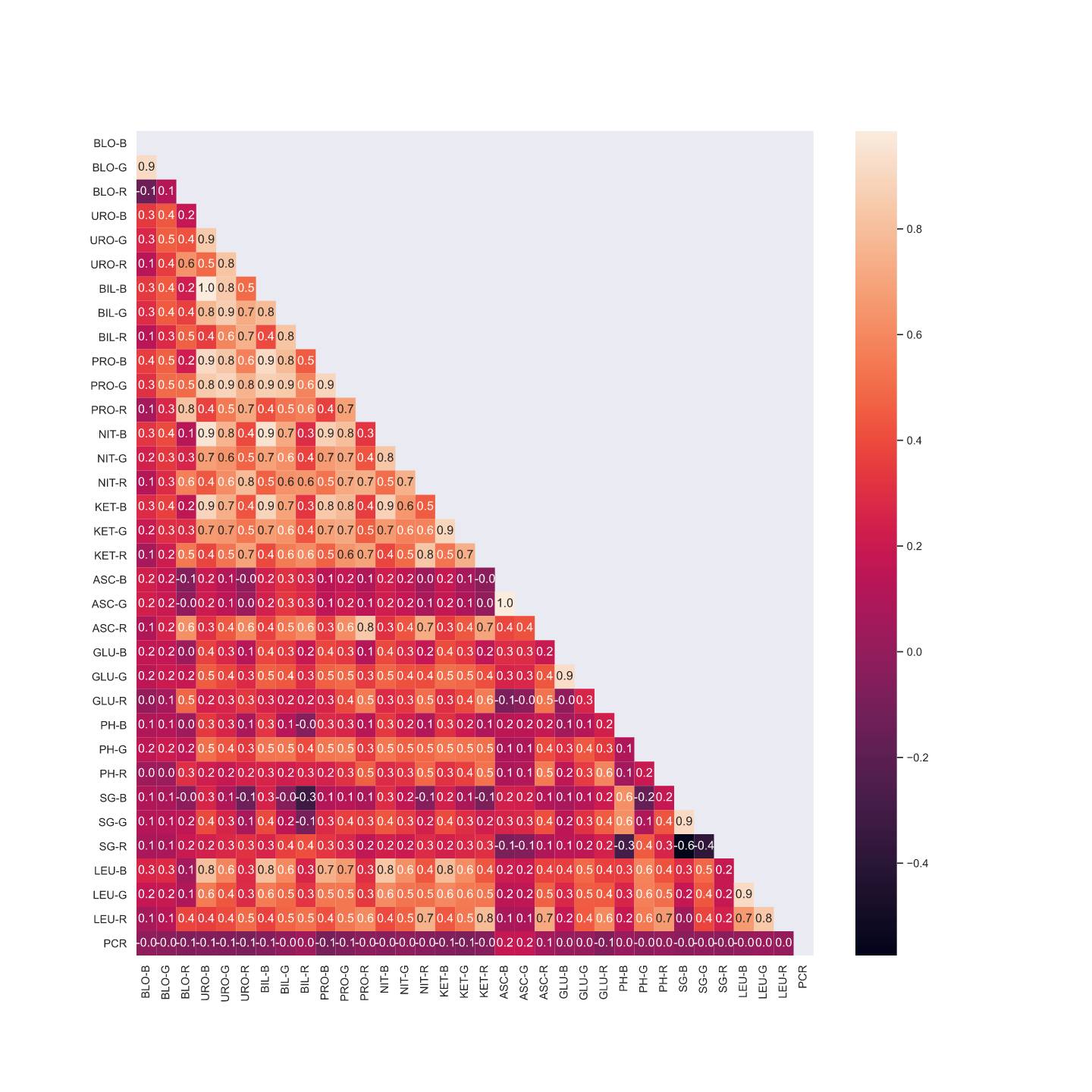}
		\caption{Correlation of urine analysis parameters}
		\label{fig:Correlation}
	\end{figure}

	For more analysis, Figure \ref{fig:Correlation} shows a heat-map of the correlation between RGB space of urine analysis features and the PCR test. In this image, dark colors indicate weaker relationships, and lighter colors indicate stronger relationships. Among the available urine analysis parameters, as a new result of this paper, ascorbic acid is more correlated with PCR test. It is considered a key parameter in the diagnosis of the disease. Also, there is a weak relationship between the presence of protein in urine and the disease of Covid-19, which was previously shown in the other studies\citep{bonetti2020urinalysis,liu2020value}.
	
\begin{figure}[H]
		\centering
		\includegraphics[scale=0.3]{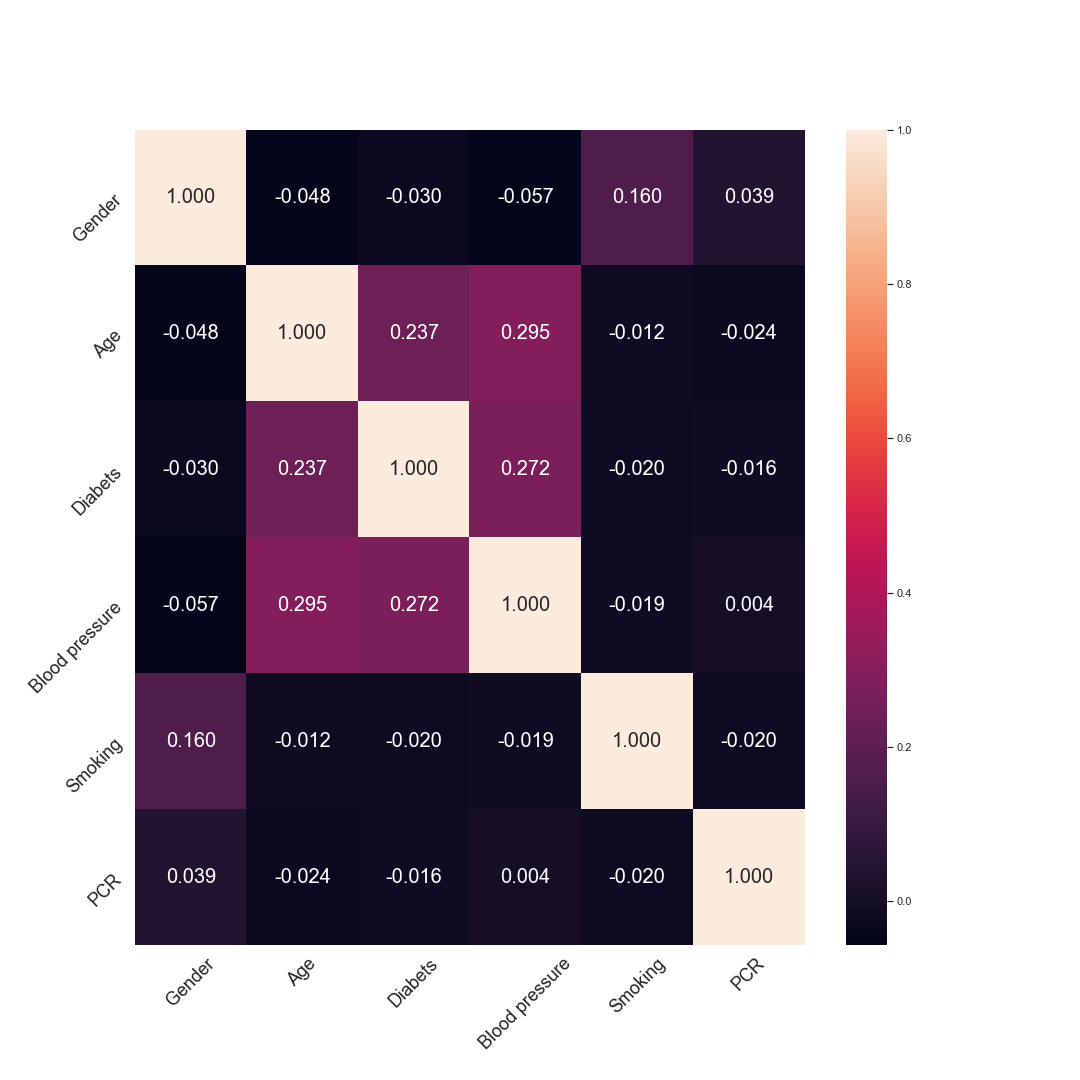}
		\caption{Correlation of clinical and PCR test values}
		\label{fig:Correlation_C}
	\end{figure}

	Figure \ref{fig:Correlation_C} shows the correlation between the clinical and PCR test variables. The first thing that can be examined in Figure \ref{fig:Correlation_C} is the weak relationship between the clinical variables and the PCR test. So results show that clinical variables are less correlated than urine parameters. This can be seen from the darkness of the last column of the image. Also, checking the image's first column shows a weak correlation between gender and smoking (0.16)\citep{abuse2020there}. Examining the second column of this image shows that age has a stronger correlation with blood pressure and diabetes variables (0.295 and 0.237)\citep{gurven2012does,suastika2012age}. The third column shows that diabetes and blood pressure variables are also correlated (0.272)\citep{bakris2004importance}.

	\subsection{Machine learning models}
	
	After going through the data collection process to develop the main part of the system, algorithms of Sickit-learn has were used. Specifically, the neural network class of this package called MLPClassifier and three different classifier namly Random Forest, Logestic Regression, and Gradient Boosting have been chosen for training. Finally, a greedy search has been engaged for setting hyperparameters of the models\footnote{Hyperparameters space provided in the supplementary materials.}. Also, for stabilization And increasing the validity of results, the mentioned process has been repeated 30 times, and the results are reported as the average of all the executions. 
	
	The importance of color spaces for image classification has been demonstrated in literature\citep{velastegui2021importance}. So in order to enrich feature representation and possibly improve the results, we have used an ensemble model on different color spaces. First, the RGB representation is converted to ten spaces including hed, hsv, lab, xyz, ycbcr, ypbpr, yuv, cie, ydbdr, yiq. Therefore, in addition to RGB, we have different color spaces for training. Then the mentioned method was used to train the model for every 10 different color spaces. The ensemble model is a selection model based on maximum votes from the trained models.
	 
\begin{figure}[H]
		\centering
		\includegraphics[scale=0.6]{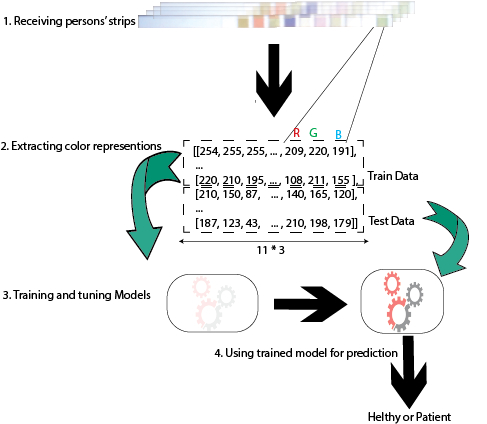}
		\caption{Steps from data collection to model training for Covid-19 screening}
		\label{fig:analysis11}
	\end{figure}

	 Figure\ref{fig:analysis11} illustrates the steps from data collection to model training for Covid-19 screening. RGB is a color model used in digital imaging and computer graphics, and it is represented by three channels: red, green, and blue. Each channel represents the intensity of a color component, and when combined, they can create a wide range of colors. To extract RGB features, we use image processing techniques to separate the three channels of an image and represent them as individual matrices. These matrices can be used as input to a machine learning algorithm to train the model. One common approach in using RGB features for machine learning is to flatten the matrices and concatenate them into a single vector, which can then be used as input to a classifier.

	\subsection{Experimental setup}
	The dataset was split into training and test sets, with 90\% of the data used for training and 10\% for testing. To optimize the performance of the models, we utilized different types of classifiers, including logistic regression, support vector machines, random forest, and neural networks.
	
	In particular, we used the MLPClassifier algorithm, which is a multi-layer perceptron algorithm that trains using backpropagation. This algorithm is available in the Scikit-learn (Sklearn) library, which is a popular machine learning toolkit in Python. Additionally, we used the GridSearchCV class in Scikit-learn for hyperparameter tuning.
	The hyperparameter settings for the MLP model included two hidden layers with (3,2) as hidden layer sizes. We repeated the process 30 times and reported the average of the measures to ensure the stability of the results.
	
	Overall, this process of training and tuning the models plays a critical role in determining the accuracy of the model in predicting the outcome variable. Using different classifiers and optimizing the hyperparameters, we can find the best model that fits training dataset and achieves the highest performance. The use of Scikit-learn and GridSearchCV makes the process more efficient and systematic, allowing for a more comprehensive search of the hyperparameter space.
	
	\subsection{Evaluation metrics}
	We use 4 measures for evaluating the models, including accuracy, precision, sensitivity and recall. Accuracy shows  the ratio of correct predictions of Covid-19 to the total number of samples and is computed as:

	\begin{align}
		Accuracy = \frac{TP + TN}{TP + TN + FN + FP}
		\label{eq:1}
	\end{align}
	\\
	\\
	Precision shows the percentages of the reported positive Covid-19 that are correctly detected:
	\begin{align}
		Precision = \frac{TP }{TP + FP}
		\label{eq:2}
	\end{align}
	\\
	Sensitivity is the probability of a positive Covid-19 test, conditioned on truly being positive. In other words, it measures the percentage of positive cases which the model identified:
	\begin{align}
		sensitivity = \frac{TP }{TP + FN}
		\label{eq:3}
	\end{align}
	\\
	Specificity is the probability of a negative Covid-19 test, conditioned on truly being negative. In other words, it measures the percentage of negative cases which the model identified or the recall of negative class:
	\begin{align}
		specificity = \frac{TN }{TN + FP}
		\label{eq:4}
	\end{align}
	\\

	Where TP represents the number of True Positive results, TN represents the number of True Negative results,  FP represents the number of False Positive results, and FN represents the number of False Negative results.

	\section{Results}
	\label{sec:Section 3}
	This section compares the average of two healthy and sick communities based on different parameters. It should be noted that for each parameter of the urine test strip, three channels of color have been formed, and a total of 33 independent T-tests have been conducted. the results are presented in Table \ref{tab:3}. It can be observed that the different values of blue, green, and red colors have a significant difference (p-value $<$ 0.05) among most cases in the studied groups. Also it can be observed that the red channel has the most difference between groups, Followed by blue, and green.

	\begin{table}[H]
		\centering
		\begin{tabular}{|c|c|c|c|c|}
			\hline
			& \textbf{Channel/Parameter} & \multicolumn{1}{c|}{\textbf{Blue}} & \multicolumn{1}{c|}{\textbf{Green}} & \textbf{Red}                 \\ \hline
			\multirow{11}{*}{\textbf{P-Value}} & Blood
			& \multicolumn{1}{c|}{0.38807}       & 0.0091                              & 0.000                            \\ \cline{2-5} 
			& Urobilinogen
			& 0.00001                                    & 0.000                                   & 0.000                            \\ \cline{2-5} 
			& Bilirubin
			& 0.00019                            & 0.62763                             & 0.00498                      \\ \cline{2-5} 
			& Protein
			& 0.000                                  & 0.000                                   & 0.06223                      \\ \cline{2-5} 
			& Nitrite
			& 0.00289                            & 0.25097                             & 0.00146                      \\ \cline{2-5} 
			& Ketone
			& 0.00001                            & 0.000                                   & 0.00116                      \\ \cline{2-5} 
			& Ascorbic Acid
			& 0.000                                  & 0.000                                   & 0.00004                      \\ \cline{2-5} 
			& Glucose
			& 0.42476                            & 0.45047                             & 0.00012                      \\ \cline{2-5} 
			& PH
			& 0.16901                            & 0.37185                             & 0.4271                       \\ \cline{2-5} 
			& Specific Gravity
			& 0.05597                            & 0.60356                             & 0.02967                      \\ \cline{2-5} 
			& Leukocytes
			& 0.99463                            & 0.04294                             & \multicolumn{1}{l|}{0.61554} \\ \hline
		\end{tabular}
		\caption{P-value of T-test for healthy and sick groups}
		\label{tab:3}
	\end{table}

	Table \ref{tab:41} shows the evaluation results of the trained model using RGB color space features with four different models. 
	Looking at the results in the table, we can see that the MLP model achieved the highest precision (50.97\%) and recall (61.45\%) scores, indicating that it had the highest proportion of true positives and correctly identified the most actual positive cases out of all the models. Additionally, the MLP model achieved the highest accuracy score (63.55\%), indicating that it had the highest proportion of correct predictions overall.

	However, it's worth noting that the Logistic Regression model achieved the highest specificity score (65.56\%), indicating that it had the highest proportion of true negatives out of all the models. This means that the Logistic Regression model was the most effective at correctly identifying negative cases.
	Overall, the table suggests that the MLP model may be the most effective at predicting positive cases, while the Logistic Regression model may be the most effective at predicting negative cases. However, the choice of model ultimately depends on the specific goals and requirements of the analysis.

	\begin{table}[H]
		\centering
		\begin{tabular}{|c|c|c|c|c|}
			\hline
			Model               & Precision      & Recall         & Specificity    & Accuracy       \\ \hline
			Random Forest       & 49.45          & 59.07          & 63.99          & 62.11          \\ \hline
			Logistic Regression & 50.91          & 60.09          & \textbf{65.56} & 63.51          \\ \hline
			Gradient Boosting   & 48.06          & 60.72          & 60.83          & 60.79          \\ \hline
			MLP                 & \textbf{50.97} & \textbf{61.45} & 64.81          & \textbf{63.55} \\ \hline
		\end{tabular}
			\caption{Evaluation of models}
	\label{tab:41}
	\end{table}
	
	Table \ref{tab:5} shows the evaluation of the ensemble model. As can be observed, the overall performance of the ensemble model is better than the model based on the RGB color space, and based on all criteria, except recall, it has improved the results.

	Based on the results of the evaluation of the ensemble model, it seems that the ensemble model is performing better than the individual models that were evaluated in the table \ref{tab:41}.

	Specifically, the ensemble model has an accuracy score of 64.69\%, which is higher than the accuracy scores of all four individual models evaluated in the previous table. Additionally, the ensemble model has a precision score of 52.63\%, which is higher than the recall scores of all the individual models. The ensemble model also has a specificity score of 66.83\%, which is higher than the specificity scores of other models.
	However, the recall score of the ensemble model is 61.13\%, which is lower than the precision scores of two of the MLP model.

	Overall, the results of the evaluation suggest that the ensemble model is a promising approach for this particular task, as it achieves higher accuracy, precision, and specificity scores than the individual models. However, the relatively low recall score is a potential concern, and further analysis would be needed to determine whether the ensemble model is the best approach for this particular problem.

\begin{figure}[H]
		\centering
		\includegraphics[scale=0.4]{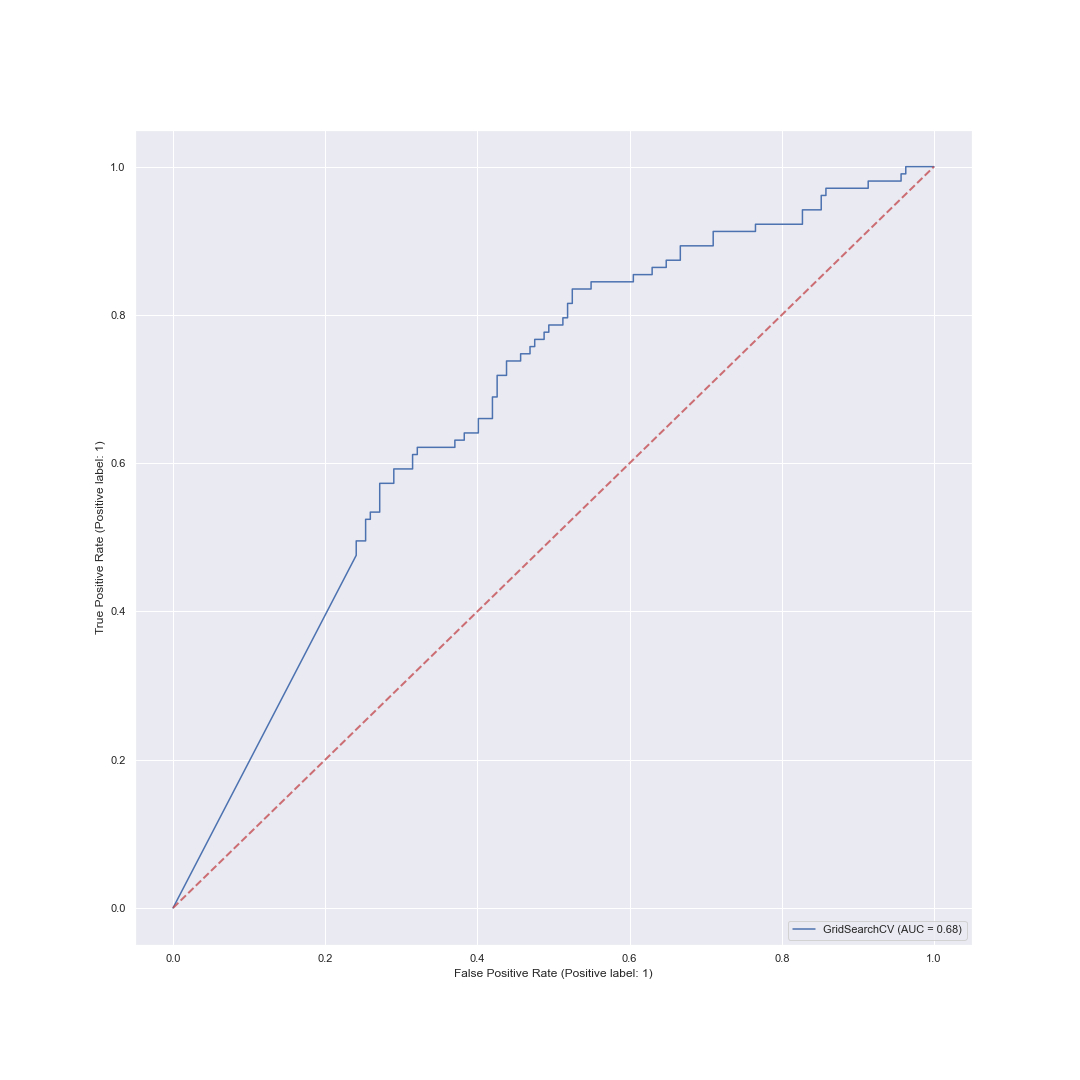}
		\caption{ROC curve of the model}
		\label{fig:ROC}
	\end{figure}

		\begin{table}[H]
		\centering
		\begin{tabular}{|c|c|c|c|c|}
			\hline
			Model               & Precision      & Recall         & Specificity    & Accuracy       \\ \hline
			Random Forest       & 49.45          & 59.07          & 63.99          & 62.11          \\ \hline
			Logistic Regression & 50.91          & 60.09          & 65.56          & 63.51          \\ \hline
			Gradient Boosting   & 48.06          & 60.72          & 60.83          & 60.79          \\ \hline
			MLP                 & 50.97          & \textbf{61.45} & 64.81          & 63.55          \\ \hline
			\textbf{Ensemble model}                 & \textbf{52.63} & 61.13          & \textbf{66.83}  & \textbf{64.69} \\ \hline
		\end{tabular}
		\caption{Evaluation of ensemble model trained on ten color spaces}
		\label{tab:5}
	\end{table}
	For determining an appropriate cut-off point of sensitivity and specificity of the test, the ROC curve can be used. The selection of the test threshold depends on the test's purpose and does not necessarily provide the same sensitivity and specificity to achieve higher accuracy \citep{hoo2017roc}. A ROC curve following the diagonal line y=x produces false positives at the same rate as true positives. Therefore, we expect the average accuracy ROC curve from the diagnostic test to lie above the y=x line (the reference line) in the upper left triangle, as shown in figure\ref{fig:ROC}.

	Due to the lack of sufficient strength of the results, a part of the space in which the model has an uncertain performance has been specified as the uncertain region, and at the time of service, if a sample from that part of the space with uncertainty is observed, it will be answered with the result of "lack of sufficient information".  In order to identify uncertain parts, the number of classifiers with positive votes between all eleven color spaces can be used. Based on this, Table \ref{tab:6} shows the results of removing the uncertain regions of space with different values. It can be observed that reducing the response space increases specificity and decreases sensitivity. Also, the growth rate of specificity is more than the sensitivity decrease. The purpose of using a screening test is to extract healthy people and send the rest to tests with higher sensitivity; for example, in table \ref{tab:6} when we have ten positive classifiers, the model has a relatively good performance Because it has a good screening and extracts about 80\% of the healthy population and only 20\% need to do other tests like the PCR.
	\begin{table}[H]
		\centering
		\begin{tabular}{|c|c|c|c|c|}
			\hline
			\textbf{\begin{tabular}[c]{@{}c@{}}The minimum number \\ of classifiers \\ with positive vote\end{tabular}} & \textbf{\begin{tabular}[c]{@{}c@{}}The ratio of \\ response space\\  to total space\end{tabular}} & \multicolumn{1}{c|}{\textbf{Accuracy}} & \multicolumn{1}{c|}{\textbf{Recall}} & \multicolumn{1}{c|}{\textbf{Specificity}} \\ \hline
			6                                                                                              & 1                                                                                              & 0.646918                               & 0.611371                             & 0.668347                                  \\ \hline
			7                                                                                              & 0.956226                                                                                       & 0.659432                               & 0.593706                             & 0.699072                                  \\ \hline
			8                                                                                              & 0.863899                                                                                       & 0.671666                               & 0.595458                             & 0.718033                                  \\ \hline
			9                                                                                              & 0.768302                                                                                       & 0.687787                               & 0.595977                             & 0.742738                                  \\ \hline
			10                                                                                             & 0.652327                                                                                       & 0.708446                               & 0.574064                             & 0.782531                                  \\ \hline
			11                                                                                             & 0.375094                                                                                       & 0.780684                               & 0                                    & 1                                         \\ \hline
		\end{tabular}
		\caption{The results of removing uncertain parts of the ensemble model}
		\label{tab:6}
	\end{table}

	\section{Conclusion}
	\label{sec:Section 5}
	
	One of the essential stages of machine learning projects is the feasibility stage, in which efforts are made to solve exploratory problems where the success or failure of the model in performing properly in that field cannot be determined with certainty. As the results of this paper and other studies\citep{morell2021urine,zolotov2022can}, due to the absence of serious distinguishing complications and the common complications caused by Covid-19 with other diseases, it is not possible to detect Covid-19 using urinary parameters with high accuracy. Nevertheless, we indicate the ability of urinary parameters like ascorbic acid to screen Covid-19 patients. Based on this, we propose an ensemble model with eleven classifiers, the model has a relatively good
	performance Because it has a good screening and extracts about 80\% of the
	healthy population and only 20\% of response space need to do other tests like the PCR.
	 For future work, efforts will be made to conduct tests after one-handed sampling (currently, samples are not taken on an empty stomach at specific times and under similar conditions), and label the samples based on the final condition of the patient to early detect whether COVID-19 will lead to severe conditions in a particular individual or not. Due to the development of the necessary interfaces for the extraction and analysis of urinary parameters, it is possible in future to find the items developed in this project to evaluate covid-19 with higher accuracy and use these parameters for screening other diseases.
	
		%

	%
	
	\section{Declarations}
	
	\subsection{Ethics approval and consent to participate}
	The research study conducted adhered to the ethical guidelines and regulations set forth by the Shiraz University of Medical Sciences. All methods employed in this study were performed in strict accordance with these guidelines to ensure the ethical treatment of participants and compliance with applicable regulations. Prior to the commencement of the study, appropriate approvals and permissions were obtained from the relevant institutional review board or ethics committee. Informed consent was obtained from all participants, and their confidentiality and privacy rights were strictly upheld throughout the study. The data collection procedures, data analysis, and reporting followed established protocols and guidelines, ensuring the integrity and validity of the findings. The researchers involved in this study were experienced professionals who were well-versed in the ethical considerations and guidelines specific to the field of study. The study findings are thus reliable and can be interpreted within the context of the ethical and regulatory framework governing research in this domain. All participants were informed about the purpose and procedures of the study and signed written informed consent forms before data collection. The participants were assured that their participation was voluntary and that they could withdraw from the study at any time without any consequences. The participants were also assured that their personal information and responses would be kept confidential and used only for research purposes.
	
	\subsection{Consent for publication}
	Consent for publication was obtained from all participants involved in this study. Participants were provided with a clear explanation of the purpose and nature of the publication, as well as any potential risks or benefits associated with it. They were assured that their personal identities and sensitive information would remain confidential and protected. Participants were given the opportunity to ask questions and were informed that their participation was voluntary, and they could withdraw their consent at any time without any negative consequences. Consent forms were signed and collected from each participant prior to their inclusion in this publication. To ensure anonymity, pseudonyms or identifiers have been used throughout the manuscript to protect the privacy and confidentiality of the participants. This study adheres to the principles outlined in the Declaration of  the Shiraz University of Medical Sciences regarding the ethical conduct of research involving human subjects. The publication of this research has been carried out with the full understanding and consent of all participants involved.

	 \subsection{Availability of data and materials}
     The current study generated and/or analysed some data that are not publicly available due to data protection and privacy reasons. These data contain personal or sensitive information that require the consent of the individuals or groups involved. The corresponding author can provide these data upon reasonable request.
	 
	 \subsection{Funding and/or Conflicts of Interests/Competing Interests}
	 The authors declare that they have no known competing financial interests or personal relationships that could have appeared to influence the work reported in this paper.

	 \subsection{Authors' contributions}
This paper was written by Mohammad Hadi Goldani as the corresponding author. Behzad Moayedi, Abdalsamad Keramatfar, Mohammad Javad Fallahi, Alborz Jahangirisisakht, Mohammad Saboori, and Leyla Badiei contributed to the research, data analysis, and manuscript preparation. Specifically, Behzad Moayedi, Abdalsamad Keramatfar and Mohammad Hadi Goldani were responsible for the implementation of the machine learning algorithms, preprocessing and evaluation of the model performance. Mohammad Javad Fallahi, Mohammad Saboori and Leyla Badiei contributed to the data collection, analysis of the experimental results and Alborz Jahangirisisakht contributed to data collection. All authors have read and approved the final manuscript and agree to be accountable for all aspects of the research, including the accuracy and integrity of the data and its analysis. Signed by, Mohammad Hadi Goldani (corresponding author) Behzad Moayedi Abdalsamad Keramatfar Mohammad Javad Fallahi Alborz Jahangirisisakht Mohammad Saboori Leyla Badiei
	
	\subsection{Acknowledgment}
 The authors would like to acknowledge the department of health affairs, Shiraz University of Medical Sciences (SUMS), Shiraz, Iran for their close collaboration and strong support in facilitating a reliable data gathering phase in this project. Also, The authors express their gratitude to the RCDAT (Research Center for Development of Advanced Technologies) for research funding.
	
	\bibliography{Covid-Project}

\end{document}